\documentclass[letterpaper,prl,twocolumn,showpacs,superscriptaddress]{revtex4}

\usepackage{graphicx}
\usepackage{hyperref}
\usepackage{bm}
\usepackage{amsmath}

\renewcommand{\section}[1]{{\par\it #1.---}}

\begin{document}

\title
{Quantum optical effective-medium theory for loss-compensated metamaterials}

\author{Ehsan Amooghorban}
\affiliation{Department of Photonics Engineering, Technical University of Denmark, DK-2800 Kgs. Lyngby, Denmark}
\affiliation{Department of Physics, Shahrekord University, Shahrekord, Iran}
\affiliation{Department of Physics, University of Isfahan, Isfahan, Iran}

\author{N. Asger Mortensen}
\affiliation{Department of Photonics Engineering, Technical University of Denmark, DK-2800 Kgs. Lyngby, Denmark}

\author{Martijn Wubs}
\email{mwubs@fotonik.dtu.dk}
\affiliation{Department of Photonics Engineering, Technical University of Denmark, DK-2800 Kgs. Lyngby, Denmark}

\date{\today}

\begin{abstract}
A central aim in metamaterial research is to engineer sub-wavelength unit cells that give rise to desired effective-medium properties and parameters, such as a negative refractive index. Ideally one can disregard the details of the unit cell and employ the effective description instead. A popular strategy to compensate for the inevitable losses in metallic components of metamaterials is to add optical gain material. Here we study the quantum optics of such loss-compensated  metamaterials at frequencies for which effective parameters can be unambiguously determined. We demonstrate that the usual effective parameters are insufficient to describe the propagation of quantum states of light.
Furthermore, we propose a quantum optical effective-medium theory instead and show that it correctly predicts the properties of the light emerging from loss-compensated metamaterials.
\end{abstract}

\pacs{42.50.Nn, 
      78.67.Pt, 
      78.20.Ci, 
      42.50.Ct 
}

\maketitle

\noindent
Metamaterials are intensely studied, since they allow the propagation and control of light in new and often counterintuitive ways. These man-made structures are composed of strongly subwavelength unit cells, with effective dielectric parameters often not occurring in nature, such as a negative refractive index~\cite{Pendry:2000a,Shalaev:2007a}. Unlike in classical optics, the possible benefits of metamaterials in quantum optics have not been explored so far, for example to manipulate single photons. More fundamentally, it is an important open question whether the same effective-medium parameters suffice to describe the propagation of quantum states of light in metamaterials.

The constituents and geometry of a unit cell can be complicated and interesting, but they are designed to allow the effective description of the metamaterial as a homogeneous medium.
For subwavelength unit cells, unique effective dielectric parameters 
can be identified, independent of the method used to retrieve them. Our results go against the common belief that experiments at the operating frequency do not reveal information about the unit cell beyond the usual effective refractive index. 

Noble metals, an important ingredient of metamaterials, are inherently lossy. For many applications it is naturally desirable to have less loss. Complementary strategies are to replace the metals by other material~\cite{Boltasseva:2011a} or to compensate for the metal loss~\cite{Ramakrishna:2003a,Nezhad:2004a,Bratkovsky:2008a,Wuestner:2010a,Fang:2010a}. As an important branch of active plasmonics~\cite{Boardman:2011a}, active loss compensation with the use of gain material has already proved experimentally successful, for example in surface plasmon polariton propagation~\cite{Leon2009,Berini:2012a} and in metamaterials~\cite{Xiao:2010a}.

In recent years the quantum optics of attenuating~\cite{Huttner:1992a,Gruner:1996a,Scheel:1998a,Artoni:1999a,Suttorp:2004a,Leonhardt:2007a,Philbin:2010a} and amplifying~\cite{Glauber:1986a,Matloob:1997a,Artoni:1998a,Vasylyev:2009a,Amooghorban:2011a} dielectric media was developed, where optical modes are described as open quantum systems.
There are important similarities with classical optics, for example the classical Green function plays a central role also in quantum optics, but new is that both with loss and gain there is quantum noise associated. Evidently, effective-medium theories that neglect quantum noise will at some point fail in quantum optics. The interesting key questions, not addressed before to our knowledge,  are whether the quantum noise of a metamaterial can be expressed in terms of effective parameters, and if so how many and how to compute them.

In this Letter we consider the propagation of quantum states of light through simple types of metamaterials, for which well-established methods to retrieve effective parameters agree very well. We show that in particular for loss-compensated metamaterials the output states can be considerably different than for their homogenized counterparts, when generalizing the usual effective-index theories to quantum optics. In response to this observation, we  propose a quantum optical effective-medium theory, with one additional effective parameter, that describes quantum light propagation in realistic metamaterials.

\section{Loss-compensated multilayer metamaterial} 
\begin{figure}[t]
\centerline{\includegraphics[width=6.5cm]{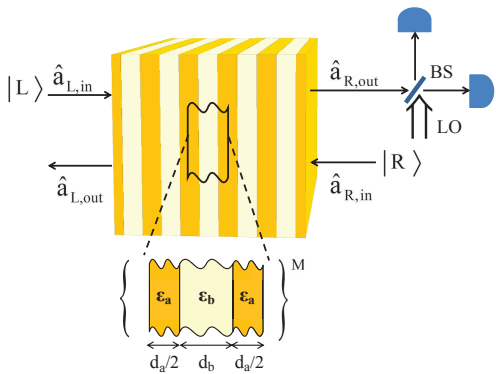}}
\caption{(Color) Sketch of the setup: Quantum states $|{L,R}\rangle$ of normally incident identically
linearly polarized light probe the metamaterial from both sides, and the output on the right is analyzed with balanced homodyne detection. The multilayer metamaterial in air has alternating  layers with  thicknesses $d_{\rm a,b}$ and volume fractions $p_{\rm a,b}=d_{\rm a,b}/d$, with $d=d_{\rm a}+d_{\rm b}$.  The two outermost layers have widths $d_{\rm a}/2$, making the metamaterial finite periodic with $M$ left-right symmetric unit cells. In Figs.~\ref{Fig:multilayer_sketch_lossgain}--\ref{Fig:Quadrature_variances_net_loss} we choose $d = 0.2 c/\omega_{0}$ and $M=5$. The susceptibilities $\varepsilon_{\rm a,b}(\omega)$ as in Eq.~(\ref{epsLorentz}) describe either loss or gain.  }\label{Fig:multilayer_sketch_lossgain}
\end{figure}

We consider loss-loss, gain-gain, and loss-gain periodic multilayers, the latter being loss-compensated metamaterials, see Fig.~\ref{Fig:multilayer_sketch_lossgain}.
Linear gain occurs when pumping the gain medium below the lasing threshold~\cite{Zyablovsky:2011a}.
For definiteness we describe the dielectric response of the layers by single-resonance Lorentz oscillator models~\cite{Matloob:1997a}
\begin{equation}\label{epsLorentz}
\varepsilon(\omega) = 1 + \left(\frac{N_{\rm down}-N_{\rm up}}{N_{\rm down}+N_{\rm up}}\right)
 \frac{\omega_{\rm p}^{2}}{(\omega_{0}^{2}-\omega^{2}-\mathrm{i}\gamma\omega)},
\end{equation}
which describe media of two-level atoms near a resonance frequency $\omega_{0}$,  with $N_{\rm up}< N_{\rm down}$ and $\varepsilon_{\rm I}=\mbox{Im}(\varepsilon)>0$ for lossy media, while gain media have inverted populations $N_{\rm up}> N_{\rm down}$ and hence $\varepsilon_{\rm I}<0$. Eq.~(\ref{epsLorentz}) satisfies Kramers--Kronig relations, and causality is essential for a consistent quantum optical description. One can parameterize the populations occurring in Eq.~(\ref{epsLorentz}) by a thermal distribution $N_{\rm th}(\omega,|T|) = [\exp(\hbar \omega/k_{\rm B}|T|)-1]^{-1}$
with an effective temperature $T$ that equals $N_{\rm up}/(N_{\rm down}-N_{\rm up})\ge 0$ for lossy and $N_{\rm down}/(N_{\rm up}-N_{\rm down})\le 0$ for amplifying media~\cite{Matloob:1997a};
$T=0\,{\mathrm{K}}$ corresponds to no excitations in the former and complete inversion in the latter.

\section{Effective parameters, two methods}
We consider nonmagnetic layers, giving a nonmagnetic effective medium with  $\mu_{\rm eff}=1$.
To determine the effective dielectric function, we compare two well-established methods: first, we take  the volume average  $\varepsilon_{\rm ave} = p_{\rm a}\varepsilon_{\rm a}+ p_{\rm b}\varepsilon_{\rm b}$, for propagation normal to the layers~\cite{Bergman:1978,Wood:2006}. As our second method, we determine the effective permittivity $\varepsilon_{\rm eff}$ from the complex amplitude reflection and transmission coefficients $r,t$ of the  metamaterial, as proposed by Smith {\em et al.}~\cite{Smith:2002a}.
We determined $\varepsilon_{\rm ave}$ and $\varepsilon_{\rm eff}$ for all structures considered in this work, and as expected since $d \ll \lambda_{0}$ for all of them, we find that they agree very well, as would other homogenization procedures~\cite{Smith:2006a,Andryieuski:2009a,Nielsen:2011a}. Thus we  denote `the' effective dielectric function by $\varepsilon_{\rm eff}$.

\section{Challenge in quantum optics}
For the propagation of quantum states of light through the metamaterial, we need to take into account that both with loss~\cite{Huttner:1992a,Gruner:1996a,Artoni:1999a,Suttorp:2004a,Leonhardt:2007a,Philbin:2010a} and with gain~\cite{Glauber:1986a,Matloob:1997a,Artoni:1998a,Vasylyev:2009a,Amooghorban:2011a} there is associated quantum noise. We do this by generalizing the
formalism by Gruner and Welsch~\cite{Gruner:1996a}  to structures with not only loss but also gain.  In this input-output formalism, the output operators are expressed as linear combinations of the input operators plus quantum noise terms $\hat F_{\rm L,R}$:
\begin{equation}\label{ainout}
\left(\begin{array}{c} \hat a_{\rm L, out} \\ \hat a_{\rm R, out} \end{array} \right) = \mathcal{A}\left(\begin{array}{c} \hat a_{\rm L, in} \\ \hat a_{\rm R, in}\end{array} \right) + \left(\begin{array}{c} \hat F_{\rm L} \\ \hat F_{\rm R} \end{array} \right).
\end{equation}
Here the $2\times 2$ matrix
$\mathcal{A}$ is the same that in classical optics connects
input and output field amplitudes.
The $\hat F_{\rm L,R}(\omega)$ on the other hand have no classical analogues, and represent linear combinations (with weights computed using the classical Green function) of the quantum noise operators associated with all layers (see below). 
The challenge for effective-medium theories in quantum optics is to predict observable effects of this quantum noise well.

\section{Homodyne detection of output light}
Let us now consider an experiment in which we probe  our metamaterial with quantum states of light, and predict the output as measured with a balanced homodyne detector~\cite{Raymer:1995a,Leonhardt:1997a}, see Fig.~\ref{Fig:multilayer_sketch_lossgain}.
As input states, we take  continuous-wave squeezed vacuum states of light~\cite{Scully:1997a}.
In quantum optics we need to take into account that the multilayer has two input ports, and to start with good squeezing, we prepare  squeezed vacuum states in both. The squeezed state $|{\rm L}\rangle = \mathcal{S}_{\rm L}|{0}\rangle$ ($|{\rm R}\rangle = \mathcal{S}_{\rm R}|{0}\rangle$) characterizes modes impinging on the left (right), with the squeezing operator
\begin{equation}\label{LRstate}
\mathcal{S}_{\rm L} = \exp\left\{\int_0^{\Delta \omega}\mbox{d}\omega\,[ \xi_{\rm L}^*\, \hat a_{\rm L, in}(\omega)\,\hat a_{\rm L, in }(2\Omega-\omega)-{\rm H.c.}]\right\}.
\end{equation}
Here, $\xi_{\rm L}=|\xi_{\rm L}|e^{\mathrm{i}\phi_{\rm L}}$ is the squeezing parameter and $\hat a_{\rm L, in}$ are  annihilation operators, see Fig.~\ref{Fig:multilayer_sketch_lossgain}. The analogous formula for $\mathcal{S}_{\rm R}$ features
$\hat a_{\rm R, in}(\omega)$ and
$\xi_{\rm R}=|\xi_{\rm R}|e^{\mathrm{i}\phi_{\rm R}}$.

The homodyne detection involves a local-oscillator field with frequency $\omega_{\rm LO}$ and phase $\phi_{\rm LO}$. The difference photocount at both detectors is proportional to the output quadrature variance~\cite{Artoni:1999a,Vasylyev:2009a}
\begin{eqnarray}\label{quadrature_variance}
\langle[ \Delta \hat E(\phi_{\rm LO},\omega_{\rm LO})]^{2}\rangle   & = & 1 +
2 \langle{\hat a_{\rm out}^{\dag}, \hat a_{\rm out}}\rangle  \\
& - & 2 \mbox{Re}[\langle{\hat a_{\rm out}^{\dag}, \hat a_{\rm out}^{\dag}}\rangle e^{2 \mathrm{i}\phi_{\rm LO}}], \nonumber
\end{eqnarray}
where we wrote $\hat a_{\rm out}$ for $\hat a_{\rm R, out}(\omega_{\rm LO})$, and use short-hand notation $\langle C, D\rangle$ for $\langle C D \rangle - \langle C\rangle\langle D\rangle$, with $\langle\ldots\rangle$ the quantum expectation value. In the time domain, the operator $\hat E(\phi_{\rm LO},t)$ is given by $\hat a_{\rm R, out}(t)\exp[\mathrm{i}(\omega_{\rm LO}t - \phi_{\rm LO}-\pi/2)] + H.c.$.
This quadrature component of the light is called squeezed if
the variance $\langle[ \Delta \hat E(\phi_{\rm LO},\omega_{\rm LO})]^{2}\rangle$ is less than unity~\cite{Blow:1990a},
the value for the vacuum state in free space.

\section{Effective-index theory in quantum optics}
In the usual classical effective-medium theories, the optical properties of a nonmagnetic metamaterial can be completely described in terms of the complex-valued effective refractive index  $\tilde n_{\rm eff} = \sqrt{\varepsilon_{\rm eff}} =  n_{\rm eff} + \mathrm{i} \kappa_{\rm eff}$ and the overall geometry (here: thickness). We call this an effective-index theory. (Not all effective-medium theories are effective-index theories, see below.)  Our first quantum optical description of metamaterials is such an effective-index theory. We adopt a very simple procedure to generalize the usual effective-index theory to quantum optics, thereby also describing quantum noise. It amounts to replacing in Eq.~(\ref{ainout}) the matrix $\mathcal{A}$ for the multilayer by another $2\times2$ matrix $\mathcal{A}_{\rm eff}$ that describes transmission and reflection of light by a single slab with thickness $M d$ and complex effective refractive index $\tilde n_{\rm eff} = n_{\rm eff} + \mathrm{i} \kappa_{\rm eff}$. Thus we have $\mathcal{A}_{11} = \exp(\mathrm{i}\omega M d/c)\mathcal{A}_{22}=r_{\rm eff}$, and $\mathcal{A}_{12} = \mathcal{A}_{21} = t_{\rm eff}$, with $t_{\rm eff}$ and $r_{\rm eff}$ the usual transmission and reflection amplitudes for such a slab. This part is the classical homogenization. Next comes the quantum homogenization, when we correspondingly replace the quantum noise operators $\hat F_{\rm L,R}$ of the multilayer by the known operators $\hat F_{{\rm L,R, eff}}$ for a single slab with refractive index $\tilde n_{\rm eff}$~\cite{Gruner:1996a}. Important for the homodyne signal is the zero-average noise operator $\hat F_{\rm eff}= \hat F_{{\rm R, eff}}(\omega)$, that depends on the effective noise-field operators $\hat \varphi_{\rm eff}(z,\omega)$ everywhere in the homogenized slab,
\begin{equation}\label{Feff}
\hat F_{\rm eff} = -\mathrm{i}\sqrt{\frac{|\varepsilon_{\rm eff, I}| \omega }{4 c}}\frac{t_{\rm eff}}{\tilde n_{\rm eff}}
\int_{0}^{d}\mbox{d}z\,f(z)\hat\varphi_{\rm eff}(z).
\end{equation}
Here, we suppressed the frequency dependence and wrote $f = f_{+}+f_{-}$ with $f_{\pm}(z,\omega) = (\tilde n_{\rm eff}\pm 1)\exp(\pm \mathrm{i}\tilde n_{\rm eff}\omega z/c)$. The commutator $[\hat\varphi_{\rm eff}(z,\omega),\hat\varphi_{\rm eff}^{\dag}(z',\omega')]$
equals ${\rm sgn}[\varepsilon_{\rm eff, I}(\omega)]\delta(z-z')\delta(\omega-\omega')$,
and thus depends on the effective medium being lossy or amplifying.

The electric-field quadrature variance Eq.~(\ref{quadrature_variance}) as measured with homodyne detection will depend on the quantum noise through the expectation value of $\langle{F_{\rm eff}^{\dag}(\omega) F_{\rm eff}(\omega')}\rangle$, which for thermal noise becomes
\begin{equation}\label{FFexpect}
\bigl\{ N_{\rm th} \theta[\varepsilon_{\rm eff, I}] - (N_{\rm th}+1)\theta[-\varepsilon_{\rm eff, I}]\bigl\}(1 - |r_{\rm eff}|^{2} - |t_{\rm eff}|^{2})\delta(\omega-\omega').
\end{equation}
Since $(1 - |r_{\rm eff}|^{2} - |t_{\rm eff}|^{2})$ is positive for net-loss and negative for net-gain media, this expectation value~(\ref{FFexpect}) is always $\ge 0$. In particular, the quantum optical effective-index theory predicts that material quantum noise vanishes at exact loss compensation, $\varepsilon_{\rm eff, I}=0$. Notice that for the limit $T=0\,{\rm K}$, Eq.~(\ref{FFexpect}) predicts $\langle F_{\rm eff}^{\dag}(\omega) F_{\rm eff}(\omega')\rangle$ to vanish for effectively lossy media, whereas the corresponding value for net gain media stays finite.

\section{Loss-loss and gain-gain multilayers}
\begin{figure}[t]
\centerline{\includegraphics[width=8.3cm]{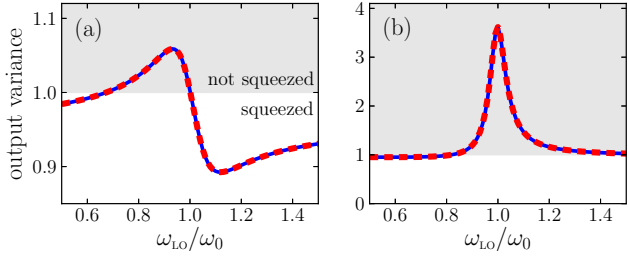}}
\caption{(Color) Output quadrature variance of Eq.~(\ref{quadrature_variance}) as measured with balanced homodyne detection, for squeezed vacuum input states with $|\xi_{\rm L}|=|\xi_{\rm R}| = 0.2$ and $\phi_{\rm L} = 2 \phi_{\rm LO}-5$ and $\phi_{\rm R} = 2 \phi_{\rm LO}-2$. Metamaterial dielectric parameters in Eq.~(\ref{epsLorentz}): $\omega_{0 {\rm a,b}}=\omega_{0}$, $\omega_{\rm p a}/\omega_{0} = 0.3$, $\gamma_{\rm a}/\omega_{0} = 0.1$, and $\omega_{\rm p b}/\omega_{0} = 0.25$ and $\gamma_{\rm b}/\omega_{0} = 0.15$. Effective temperatures $T_{\rm a}=T_{\rm b}=0$. Panel (a): loss-loss   multilayer. (b): gain-gain multilayer. Red dashed lines (quantum optical effective-index theory) overlap blue solid lines (exact multilayer theory).}\label{Fig:LLGG}
\end{figure}

As a first test of the quantum optical effective-index theory, we present in Fig.~\ref{Fig:LLGG} the output quadrature variance for loss-loss and gain-gain metamaterials, all at $T=0$.
As the figure shows, the theory predicts the homodyne signal well, for both multilayers. For the loss-loss multilayer at zero temperature quantum noise can be neglected so the agreement only confirms that $\mathcal{A}$ can be replaced by $\mathcal{A}_{\rm eff}$, as we know from classical optics. By contrast, for gain-gain multilayers, the effects of quantum noise in the output cannot be neglected at zero temperature, but the figure illustrates that here also the quantum optical effective-index theory is accurate.
The same holds true (not shown) for the structures of Fig.~\ref{Fig:LLGG} at 
finite temperatures $T_{\rm a}=T_{\rm b}\ne 0$.
However, if due to pumping the effective temperatures $T_{\rm a}$ and $T_{\rm b}$ are different in both layer types, then the quantum optical effective-index theory does not specify what distribution should be used in Eq.~(\ref{FFexpect}).

\section{Breakdown of quantum optical effective-index theory}
\begin{figure}[t]
\centerline{\includegraphics[width=8.3cm]{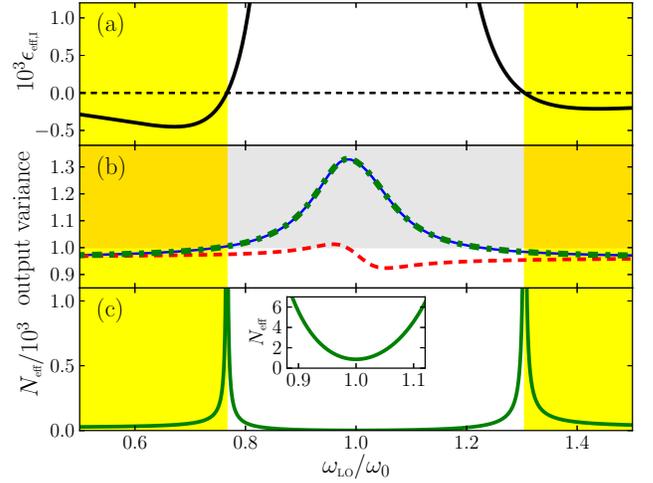}}
\caption{(Color) Loss-compensated metamaterial with geometry of Fig.~\ref{Fig:multilayer_sketch_lossgain}, with lossy $\rm a$-layers and amplifying $\rm b$-layers. Material parameters and input states as in Fig.~\ref{Fig:LLGG}.  Panel~(a): frequency dependence of $\varepsilon_{\rm eff, I}$, showing net loss around $\omega_{0}$ and net gain (yellow) elsewhere. Exact loss compensation occurs at $0.77\omega_{0}$ and $1.30\omega_{0}$. (b): Output quadrature variances predicted with the quantum optical effective-index theory (red dashed line), and with the quantum optical effective medium (QOEM) theory (dash-dotted green line); the latter overlaps the exact multilayer calculation (solid blue line). (c): corresponding effective noise photon distribution $N_{\rm eff}$ of Eq.~(\ref{Neffalphaeff}).}\label{Fig:Quadrature_variances_net_loss}
\end{figure}

For loss-compensated metamaterials, we see in Fig.~\ref{Fig:Quadrature_variances_net_loss}(b)
already for $T_{\rm a}= T_{\rm b}=0$ a clear failure  of the quantum optical effective-index theory,
which predicts the output light to be squeezed
for almost all values of $\omega_{\rm LO}$ near the material resonance $\omega_{0}$, whereas the full multilayer calculation shows that actually the output state will not at all be squeezed in the whole range of $\omega_{\rm LO}/\omega_{0}$ between 0.76 and 1.21. The output of a loss-compensated material is noisier than of a homogeneous slab with the same $\tilde n_{\rm eff}$. 

This breakdown of effective-index theory does not depend on the simple response model of Eq.~(\ref{epsLorentz}). In the Supplement~\cite{Supplement} we show the same breakdown and its experimental relevance for multilayers of silver and a pumped dye medium, using measured dielectric functions.

\section{Quantum-optical effective medium theory}
We will now derive a quantum-optical effective medium (QOEM) theory,  that does give accurate predictions also for loss-compensated metamaterials. In contrast to the previous one, it is not an effective-index theory. Our approach is to distill solely from a unit cell
not only the usual $\varepsilon_{\rm eff}(\omega)$, but also an effective noise photon distribution $N_{\rm eff}(\omega)$.

Applying the general multilayer theory to one unit cell only, we compute the noise contribution  $\langle{F_{\rm unit}^{\dag}(\omega) F_{\rm unit}(\omega')}\rangle$ to the
output variance. The use of the general theory ensures that we do not introduce the unphysical noise compensation that we found above to be inherent in the quantum optical effective-index description. Since the unit cell is much smaller than an optical wavelength, it suffices to Taylor expand
the average to first order in the layer thicknesses $d_{\rm a,b}$. Analogous to Eq.~(\ref{FFexpect}), we equate $\langle{F_{\rm unit}^{\dag}(\omega) F_{\rm unit}(\omega')}\rangle$ with $\bigl\{ N_{\rm eff} \theta[\varepsilon_{\rm eff, I}] - (N_{\rm eff}+1)\theta[-\varepsilon_{\rm eff, I}]\bigl\}(1 - |r_{\rm unit, eff}|^{2} - |t_{\rm unit, eff}|^{2})\delta(\omega-\omega')$, the product of the yet unknown $N_{\rm eff}$ and the net gain or loss of a unit cell, the latter computed in the usual classical effective-index theory
and also to first order in the thickness. Solving for $N_{\rm eff}$, we obtain as a main result the effective noise photon distribution
\begin{equation}\label{Neffalphaeff}
N_{\rm eff}(\omega)  = -\frac{1}{2}+\frac{1}{2}\sum_{j={\rm a,b}}\zeta_{j}[2 N_{\rm th}(\omega,|T_{j}|)+1],
\end{equation}
in terms of the dielectric parameters of the unit cell
\begin{equation}
\zeta_{j}  =  p_{j}\left|\frac{\varepsilon_{j, {\rm  I}}}
{\varepsilon_{\rm eff, I}}\right| \label{zetaj}.
\end{equation}
Thus, our  QOEM theory predicts that the quantum noise contribution $\langle F_{\rm eff}^{\dag}(\omega) F_{\rm eff}(\omega')\rangle$ to the output variance of a multilayer metamaterial consisting of arbitrarily many unit cells is given by Eq.~(\ref{FFexpect}), but with the thermal distribution $N_{\rm th}$ replaced by the effective distribution $N_{\rm eff}$ of Eq.~(\ref{Neffalphaeff}).  Fig.~\ref{Fig:Quadrature_variances_net_loss}(b)
illustrates the very good agreement between the output variances
computed with the exact theory and with the QOEM theory for a loss-compensated metamaterial. 
Calculations for similar structures in our Supplement support this main result and its experimental relevance~\cite{Supplement}.
$N_{\rm eff}$ grows when the loss is more exactly compensated by gain [smaller $|\varepsilon_{\rm eff, I}|$, see Fig.~\ref{Fig:Quadrature_variances_net_loss}(a)] or when the same value $|\varepsilon_{\rm eff, I}|$ results from compensating more loss with more gain (larger $|\varepsilon_{j,{\rm I}}|$). When loss is exactly compensated $N_{\rm eff}$ even diverges, while the output variance
stays finite, compare Figs.~\ref{Fig:Quadrature_variances_net_loss}(a,b,c).

The amount by which the sum $\sum_{j={\rm a,b}}\zeta_{j}$ exceeds unity is a measure of how much loss is compensated by how much gain, which is information beyond $\varepsilon_{\rm eff}(\omega)$.
For loss-loss and gain-gain metamaterials
the $\zeta_{\rm a}$ and $\zeta_{\rm b}$ add up to unity. Only in case $T_{\rm a}=T_{\rm b}$, as for a passive lossy metamaterial at room temperature, we then find that $N_{\rm eff}$ reduces to the thermal distribution $N_{\rm th}$. Thus our QOEM theory explains our earlier observation that the quantum optical effective-index theory worked well in Fig.~\ref{Fig:LLGG}.

\section{Conclusions and outlook} 
Quantum optics of metamaterials is a new research topic, for which we propose two effective-medium theories. The first and simplest one is an effective-index theory, involving only the effective index as the effective parameter, just like in classical optics. This quantum optical  effective-index  theory is valid for metamaterials in thermal equilibrium, but otherwise typically breaks down. Unlike in classical optics, the effective dielectric function $\varepsilon_{\rm eff}$ does not always suffice to describe the output even of perfectly homogenizable metamaterials.
In particular for loss-compensated metamaterials, we showed that balanced homodyne detection can be used to `open the effective-index blackbox' and to detect in which of two metamaterials with the same $\tilde n_{\rm eff}$ most loss compensation occurs, because loss compensation is not conveniently accompanied by quantum noise compensation. We showed that on this point the quantum optical effective-index theory fails, and also fails without loss compensation when different effective temperatures coexist due to pumping. Indeed this is quantum electrodynamics out of equilibrium~\cite{Antezza:2008a,Krueger:2011a}.

Moreover, we proposed and tested a quantum-optical effective medium (QOEM) theory that works for loss-compensated and for gain-gain and loss-loss metamaterials alike. The effective dielectric function is the usual one, but it is complemented by an effective noise distribution $N_{\rm eff}(\omega)$  which involves dielectric properties beyond $\varepsilon_{\rm eff}$, namely the positive loss and gain parameters $\zeta_{\rm a,b}$.
These central results of Eqs.~(\ref{Neffalphaeff},\ref{zetaj}) can be generalized without difficulty to more than two layers per unit cell or continuously varying $\varepsilon(z)$. The formulae do not depend explicitly on the unit cell being layered, which suggests their validity for a much larger class of metamaterials.
This deserves further study, but for a slab of loss-compensated 3D fishnet metamaterial~\cite{Wuestner:2010a,Fang:2010a}, one could again calculate the quantum noise, this time using Green-function methods~\cite{Scheel:1998a,Suttorp:2004a,Philbin:2010a}, and again average.

Our findings are not just relevant for homodyne detection of squeezed light, but for the optics of metamaterials in general, for example when employing  superlenses based on loss-compensated multilayers as proposed in Ref.~\onlinecite{Ramakrishna:2003a} to focus single- or few-photon pulses of light. The quantum noise grows with thickness of the metamaterial and is independent of the optical input states, and thus is relatively more important for weaker input fields (fewer photons) and negligible for bright light. But whatever input state is used, there will be additional output variance due to the quantum noise in pumped metamaterials, underestimated by effective-index theories, but well predicted by our QOEM theory. Analogous to metamaterials, it would be interesting to study the quantum optics of loss-compensated slow-light media~\cite{Grgic:2012a}.

In summary, we identified a fundamental relation between loss compensation and quantum noise, constituting a breakdown of standard effective-index theory, that should be taken into account for few-photon metamaterial applications. We expect our quantum optical effective-medium theory to become a useful tool for the design and use of active metamaterials.

\begin{acknowledgments}
E.A. thanks the Iranian Ministry of Science, Research, and Technology for financially supporting his stay at the Technical University of Denmark.
\end{acknowledgments}

\end{document}